\documentclass[aps,preprint]{revtex4}
\usepackage{amsmath,epsfig,graphicx,amssymb}

\begin{document}

\title{Exploration of  electronic quadrupole states
  in atomic clusters by two-photon processes}

\author{V.O. Nesterenko$^1$, P.-G. Reinhard$^2$, T. Halfmann$^3$,
and E. Suraud$^4$}
\date{\today}
\affiliation{$^{1}$Laboratory of Theoretical Physics, Joint Institute for
Nuclear Research, Dubna, Moscow region, 141980, Russia}
\email{nester@theor.jinr.ru} \affiliation{$^{2}$ Institute of Theoretical
Physics II, University of Erlangen-Nurnberg, D-91058, Erlangen, Germany}
\affiliation{$^{3}$ Department of Physics, Technical University Kaiserslautern,
D-67653, Germany} \affiliation{$^{4}$ Institute of Electronics, Bulgarian
Academy of Sciences, Sophia, Bulgaria}
\affiliation{$^{4}$ Laboratoire de Physique Quantique, Universit{\'e} Paul Sabatier,
 118 Route de Narbonne, 31062 cedex,  Toulouse, France}

\begin{abstract}
We analyze particular two-photon processes as possible means to explore electronic quadrupole
states in free small deformed atomic clusters. The analysis is done
in the  time-dependent local density approximation (TDLDA). It is shown that
the direct two-photon population (DTP) and off-resonant stimulated Raman (ORSR)
scattering can be effectively used for excitation of
the quadrupole states in high-frequency
(quadrupole plasmon) and low-frequency (infrared) regions, respectively.
In ORSR, isolated dipole particle-hole states as well as the tail
of the dipole plasmon can serve as an intermediate state. A simultaneous study
of low- and high-frequency quadrupoles, combining DTP and ORSR, is most
effective. Femtosecond pulses with intensities $I = 2\cdot 10^{10} - 2\cdot
10^{11} W/cm^2$ and pulse durations $T = 200 - 500$ fs are found to be optimal.
Since the low-lying quadrupole states are dominated by one single electron-hole
pair, their energies, being combined with the photoelectron data for hole
states, allow to get the electron spectrum above the Fermi level and thus
greatly extend our knowledge on the single particle spectra of clusters.
Besides, the developed schemes allow to estimate the lifetime of the quadrupole
states.
\end{abstract}
\maketitle

\section{Introduction}

In recent years,  the achievements in laser technologies have lead to a
remarkable progress in the analysis of electronic degrees of freedom in
atomic clusters, for an extensive review see \cite{Reinbook}.
For example, photoelectron spectra (PES) can now be measured with a high
accuracy and for a broad variety of clusters
\cite{Wrigge,Moseler,Alum,semi,SIC_rost,c60_cam,c60_cam2}.
However, a wealth of electronic modes still remains unexplored.
In clusters with the diameter far below the laser wavelength (i.e. in clusters with
the number of atoms $N < 10^6 - 10^8$), the laser light couples only to dipole
($\lambda =1$) states. Hence we gain the well known dipole plasmon.
At the same time, it is still very hard to access the electronic modes with
higher multipolarity $\lambda  > 1$.  Multi-photon processes can generally
give a way to these modes. For example, two photons couple to quadrupole
($\lambda =2$) modes \cite{Ne_PRA_2004}. Just this case,
investigation of quadrupole modes
of valence electrons in two-photon processes, will be scrutinized in the
present paper. In a  previous publication \cite{Ne_PRA_2006}, we have studied
the scenario of two-photon processes where both photons originate from
the same laser
pulse and so have the same frequency. In this paper, we will analyze an
alternative two-photon technique employing two different laser
frequencies. It will be
shown that the combined implementation of both techniques
is most optimal for the exploration of electronic quadrupole states.

A particularly interesting aspect emerges for low-frequency (infrared) quadrupole
modes in small deformed clusters.
It was found that these modes are dominated by a single electron-hole (1eh) pair
\cite{Ne_PRA_2004,Ne_PRA_2006} and their spectra are close to the pure
1eh energy
differences $\epsilon_{eh}= \epsilon_{e}-\epsilon_{h}$.
As a result, measuring the energy
$\epsilon_{eh}$ in a two-photon process and the energy of the hole (occupied) state
$e_h$ by PES \cite{Wrigge,Moseler,Alum}, we gain information on the
particle energy $\epsilon_{e}$. This allows to determine the full single-particle
spectrum of valence electrons  near the HOMO-LUMO (highest occupied
molecular orbital - lowest unoccupied molecular orbital) gap.  Being sensitive
to diverse cluster's features (equilibrium shape, ionic structure, ...), the
electronic spectra can in turn serve for investigations of these features. Besides they
constitute a critical test of any theoretical description.

It is worth noting that,
unlike the dipole states with their high frequencies and strong collective mixing,
the quadrupole states of interest mainly originate from the deformation
splitting \cite{Ne_PRA_2004,Ne_PRA_2006}.  Their energy scale is thus quite
small and they usually lie in the infrared region $< 1 eV$. In small clusters
the spectrum of these states is very dilute. This prevents collective mixing
of the states, favors their $1eh$ nature, and simplifies the
experimental discrimination. Different kinds of small deformed clusters
(free, supported, embedded) can be explored for the infrared quadrupole modes.
In this paper we will consider the simplest case of free clusters.

Two-photon processes allow to excite not only the low-frequency quadrupole
modes, but also high-frequency quadrupole states in the regime of the
quadrupole plasmon. In fact, these plasmon states carry a large quadrupole
strength and thus rather easily respond to two-photons probes. The two sorts of
quadrupole modes can be characterized in terms of transitions between major
quantum shells of the cluster mean field \cite{Ne_PRA_2004}.
The low-frequency quadrupole modes correspond to $1eh$ excitations within the
valence quantum shell $N$ ($\Delta N =0$ modes). The high-frequency states in
the quadrupole-plasmon range correspond to the excitations over two major
shells ($\Delta N =2$ modes). There are still no any experimental
data about both kinds of modes. But their investigation could
deliver a valuable spectroscopic information.

\begin{figure}
\centerline{
\includegraphics[height=10cm,width=3cm,angle=-90]{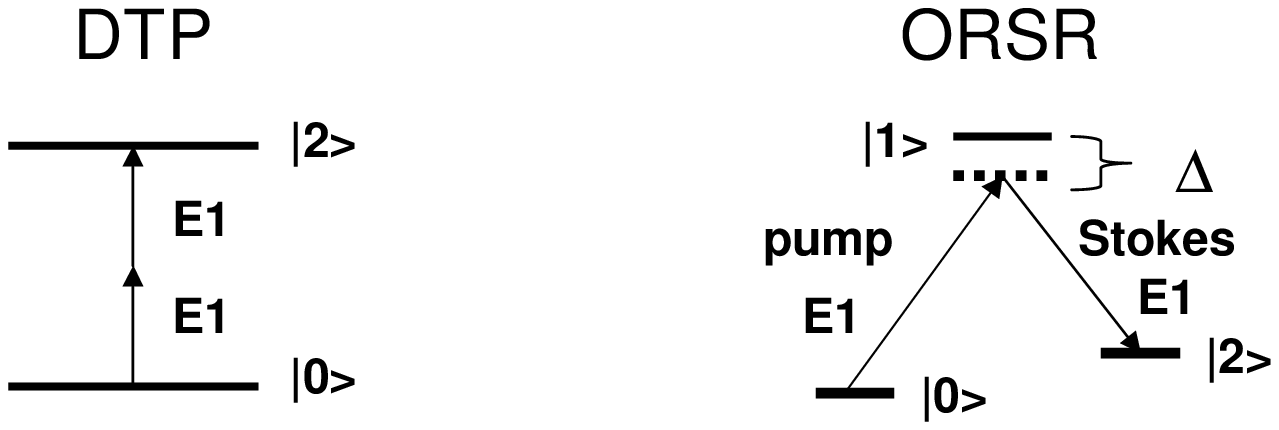}
}
\caption{\label{fig:lam_sys_fig1}
Schemes of two-photon processes: direct two-photon (DTP) in a two-level system
and off-resonant stimulated Raman (ORSR) in a three-level $\Lambda$-system.
The initial $|0\rangle$, intermediate $|1\rangle$, and target $|2\rangle$ states have
the orbital moments $\lambda=$0, 1, and 2, respectively.
In ORSR the pump dipole pulse
couples the ground and intermediate states while the Stokes dipole pulse
provides the coupling of the intermediate and target states.
$\Delta$ is the detuning from the intermediate dipole state $|1\rangle$.
The purpose of both DTP and ORSR is the population of the
target quadrupole state $|2\rangle$.
}
\end{figure}

As mentioned above, the excitation of quadrupole states needs at least
a two-photon process. A variety of such processes is known in atomic and
molecular spectroscopy \cite{Scoles,SEP,Vitanov,Berg}. However, to our knowledge,
none has been applied so far in experimental investigations of atomic
clusters. Some of these processes, namely the direct two-photon population (DTP)
\cite{Ne_PRA_2006}, the off-resonant stimulated Raman scattering (ORSR)
\cite{SEP}, and the stimulated Raman adiabatic passage (STIRAP)
\cite{Vitanov,Berg}, seem to be quite promising \cite{NY} and are thus worth a
closer inspection. As is shown below, some particular cluster properties, e.g.
high probability of undesirable plasmon population, complicates
implementation of two-photon schemes. Hence we need a detailed analysis
based on realistic calculations.
As a first step in this direction, the pump-probe DTP method was recently
investigated \cite{Ne_PRA_2006}. In this method the electronic infrared
quadrupole state is generated via the direct resonant two-photon (two-dipole)
excitation by the pump laser, see the DTP scheme at the left part of
figure \ref{fig:lam_sys_fig1}. The population of the quadrupole state
is detected through the
appearance of satellites in the photoelectron spectra produced by a probe
pulse (not shown in figure \ref{fig:lam_sys_fig1}).  Femtosecond pump and probe pulses with
intensities $I = 2\cdot 10^{10} - 2\cdot 10^{11} W/cm^2$ and pulse duration $T
= 200 - 500$ fs were found to be optimal.  The systematic  TDLDA
calculations have shown that the method is very robust and delivers not only
the $1eh$ spectrum but also the lifetime of the $1eh$ pairs.

In the present paper, we aim to inspect an alternative two-photon method,
off-resonant stimulated Raman scattering (ORSR).  In this method, the target
quadrupole state is populated by two different dipole transitions via an
intermediate dipole state.  Hence we deal with the so called $\Lambda$-system, see
right part of figure \ref{fig:lam_sys_fig1}.  The pump pulse provides the coupling of the initial
(ground) state $|0\rangle$ to the intermediate state $|1\rangle$.  The Stokes pulse
couples $|1\rangle$ with the quadrupole target state $|2\rangle$, altogether stimulating
the transition to the target state. The pulses have to maintain the two-photon
resonance condition $\omega_p - \omega_s = \omega_2$, i.e. the
difference of the pump and Stokes frequencies must coincide with the frequency of the
target quadrupole state.  Isolated dipole states of $1eh$ nature as well as
the dipole plasmon can serve as the intermediate state $|1\rangle$. However, one
should avoid a real population of the intermediate state to prevent
undesirable leaking into competing channels. This is especially important for
the dipole plasmon which decays via a fast Landau damping associated
with a short lifetime ($\sim 10-20$ fs) \cite{Reinbook}. To avoid the actual
population of $|1\rangle$, we will use an appreciable detuning $\Delta$ from the
energy of this state, hence the reference to {\it off-resonant} process. A
considerable detuning is the crucial point in our scheme. Detection of the
population of the target
quadrupole states (both at low- and high-frequency) in the ORSR can be done by
a probe pulse in the same way as in the DTP \cite{Ne_PRA_2006}.
As compared with DTP, the ORSR scheme is more involved since it requires not two
but three different pulses (pump, Stokes and probe). At the same time, the
ORSR allows to explore the low-lying quadrupole states by lasers in the region
of visible light and hence is an interesting alternative to DTP. Besides,
ORSR is widely used in atomic and molecular physics. It is certainly worth to
assay this method for atomic clusters as well.

In the present study, we will apply the ORSR to low-frequency quadrupole
states. However, as is shown below, it is hard to explore these states
without touching the high-frequency quadrupoles which very easily respond
to any two-photon probes. Both kinds of quadrupole states should thus be
involved into a realistic exploration scheme. The high-frequency quadrupoles
can hardly be studied within the ORSR in $\Lambda$-configuration because in
this case we would need the intermediate dipole state lying above the
quadrupole plasmon \cite{comm1}.
It is hence better to explore these
quadrupoles by DTP. As a result, we naturally come to a combined analysis
implementing the ORSR for the low-frequency quadrupoles and the DTP for
their high-frequency counterparts. The aim of the present paper is to develop
optimal schemes for the combined DTP-ORSR method.

The paper is outlined as follows. In Section
2 the calculation scheme is sketched. In Sec. 3 the ORSR excitation of the
low-frequency quadrupole is discussed for two cases of the intermediate state:
an isolated dipole state and tail of the dipole plasmon. Stability of the process
to variation of the main parameters is scrutinized. The DTP population of the
quadrupole plasmon states is outlined and the general scheme for the combined
DTP-ORSR experiment is developed.  The conclusions are drawn in Sec. 4.

\section{Calculation scheme and test case}

The theoretical and numerical framework of our study are explained in detail
elsewhere \cite{Reinbook,Cal}. So we summarize here only the key points.  The
calculations are performed within the time-dependent local density
approximation (TDLDA) using the exchange-correlation functional of \cite{ex_corr}
and an averaged self-interaction correction \cite{adsic}.  As a first step, the static
single-electron wave functions $\bar\phi_i (\vec r)$ are calculated from the
stationary Kohn-Sham equations. Then time evolution of the single-electron
wave functions $\phi_i (\vec r,t)$ is computed starting from the initial
condition $\phi_i (\vec r,t=0)=\bar\phi_i (\vec r)$.  The ionic background is
treated in the soft jellium approximation \cite{Reinbook}.
This approximation, although a bit daring, is capable of reproducing the basic
trends of shapes and subsequent plasmon spectra of Na clusters
\cite{Brack,Hee93,Mon95b}. In the present paper we consider axially deformed
clusters and therefrom Na$_{11}^+$ as a particularly suitable test case.
The axial symmetry and jellium approximation together greatly reduce the computational
effort and thus allow huge scans in the multi-parameter space of multi-photon
processes even in deformed clusters. Absorbing boundary conditions are employed
for the description of photoionization. The numerical handling is performed by
standard methods (gradient iterations for the ground state, time splitting for time
propagation). The excitation spectra in the linear response regime are computed
in TDLDA by standard techniques of spectral analysis \cite{Cal,spectran}. The
laser induced dynamics is simulated by adding to the TDLDA the laser pulses as
classical external dipole fields of the form $W(t)=E_0\,z\,\sin^2(\pi t/T)\cos
(\omega t)$ with the field strength $E_0$ ($\propto$ square root of the intensity
$I$) lasting for one half-cycle of the profile function $\sin^2(\pi t/T)$. The field
is applied in $z$-direction (the symmetry axis of the system);
$\omega$ is the frequency and $T$ is the pulse duration.
\begin{table}[t]
\begin{center}
\caption{\label{tab:spectrum}
The spectrum of the quadrupole states below 1 eV in Na$_{11}^+$,
approximated by the energies of the dominant $1eh$ pairs. The structure of
$1eh$ pairs is done in terms of the Nilsson-Clemenger quantum numbers
$[Nn_z\Lambda]$ \protect\cite{Clem}. See text for more details.
}
\begin{tabular}{ccc}
\hline
$\lambda\mu$ & $\hbar\omega $ [eV] & $[Nn_z\Lambda]_e \; [Nn_z\Lambda]_h$ \\
\hline
21 & 0.41 & [211] [220]\\
22 & 0.60 & [202] [220]\\
20 & 0.75 & [200] [220]\\
\hline
\end{tabular}
\end{center}
\end{table}
Small deformed sodium clusters are optimal for a first exploratory analysis.  We
consider here the test case Na$_{11}^+$.  It is strongly prolate which
provides a comfortably strong collective splitting of the plasmon
resonance and of the single-electron spectrum. Its infrared spectrum below 1
eV is very dilute and displays only three well separated electronic
levels, namely the quadrupole modes of multipolarity $\lambda\mu$=20,
21 and 22 (see Table 1). Following our estimations \cite{Ne_PRA_2004},
these modes almost correspond to pure $1eh$ states as indicated in
Table 1. The collective shifts of these states through the Coulomb
residual interaction are modest, e.g.  $\sim 0.05$ eV for
$\lambda\mu$=20, which corroborates their $1eh$ structure.  This
feature is a direct consequence of the dilute spectrum, because large
energy intervals between the levels prevent their collective mixture.
In Na$_{11}^+$ dipole states start above 1 eV and so are well
separated from the low-frequency quadrupole modes. Such a spectral
separation is a general feature of small deformed clusters. As was
mentioned above and as can be seen from Table 1, the low-frequency
quadrupole modes are represented by the $1eh$ transitions inside the
valence shell ($\Delta N =0$). Moreover, most of these
modes arise due to the deformation splitting of the levels and so
their energy scale is small.  Instead, the dipole modes are generated
by $\Delta N =1$ electron-hole transitions between the neighbor quantum
shells and thus acquire much higher energies.  The effective energy
separation of the quadrupole and dipole modes favors discrimination of
the low-frequency quadrupole spectrum.

%
\begin{figure}
\centerline{
\includegraphics[height=9cm,width=6.5cm,angle=-90]{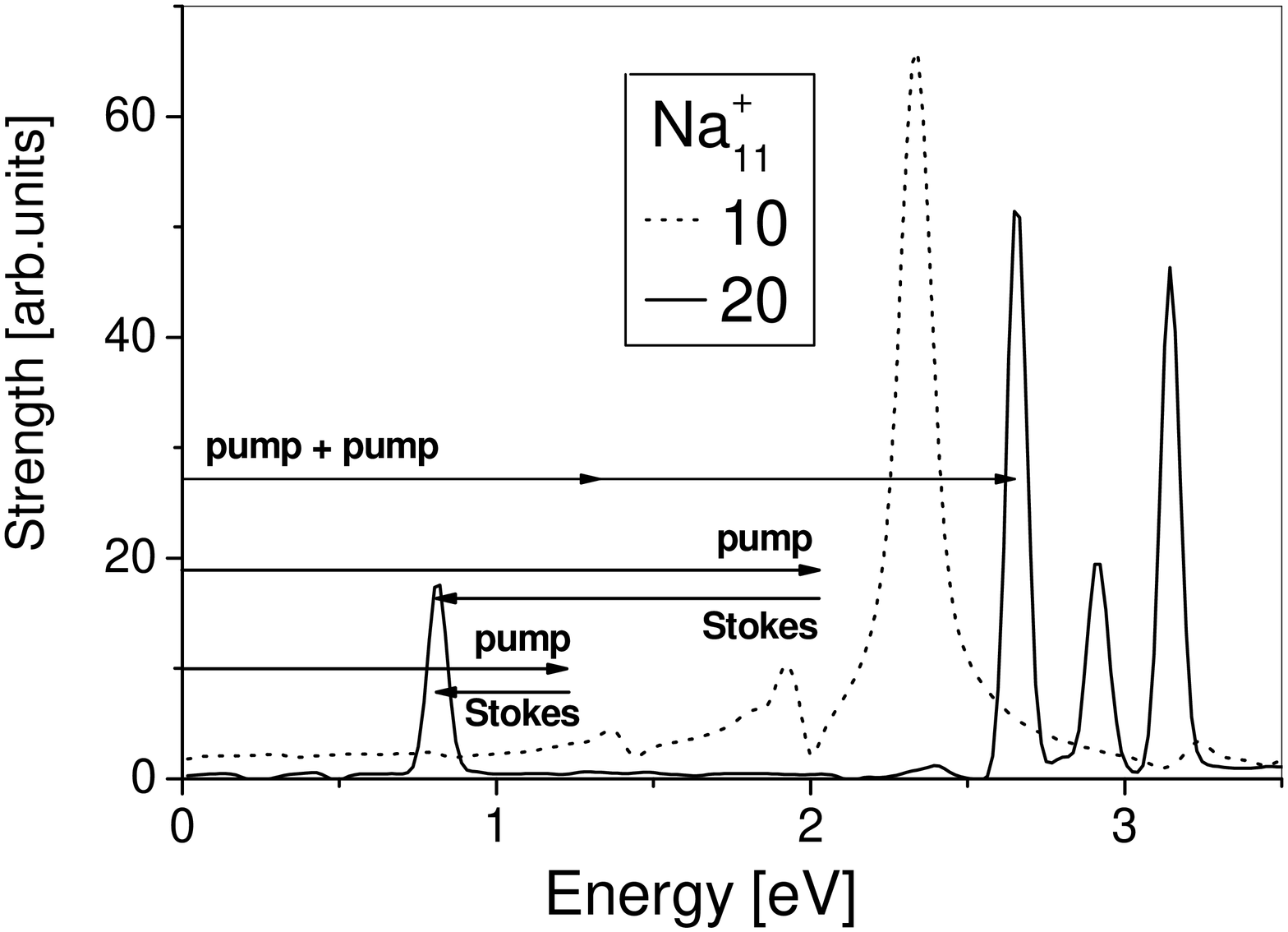}
 }
\caption{\label{fig:be12_fig2}
Dipole ($\lambda\mu =$10) and quadrupole ($\lambda\mu =$20)
strength distributions in Na$^+_{11}$. The large peaks above 2 eV represent
the states of the dipole and quadrupole plasmons. The sets of
horizontal arrows depict
two-photon (two-dipole) processes: DTP (pump + pump) and ORSR (pump + Stokes).
The latter runs via the isolated dipole state at 1.35 eV and the tail of the
dipole plasmon.  The low- and high-frequency quadrupole states of interest
are seen as peaks at 0.8 and 2.6 eV.
}
\end{figure}
In what follows, we will concentrate on the states with
$\lambda\mu$=20.  This suffices for our exploration. Besides, the limitation
to $\lambda\mu$=20 allows to maintain the axial symmetry which, in turn,
reduces computational expense.  Figure \ref{fig:be12_fig2} shows the relevant part of
the excitation spectrum in terms of the dipole ($\lambda\mu=10$) and
quadrupole ($\lambda\mu=20$) photo-absorption strengths.  The low- and
high-frequency quadrupole modes of interest are seen at the energies
$e_{20}=$0.8 and 2.6 eV.  The horizontal arrows depict the DTP and
ORSR processes considered below.  Two ORSR versions are discussed. In
the first case, the process runs via the isolated dipole state at 1.35
eV. In the second case, it proceeds via the intermediate region
between the isolated dipole state at 1.9 eV and the dipole plasmon. In
both cases, there is an appreciable detuning from the dipole states
such that one deals only with remote tails of the states.
%
\begin{figure}
\centerline{
\includegraphics[height=11cm,width=8cm,angle=-90]{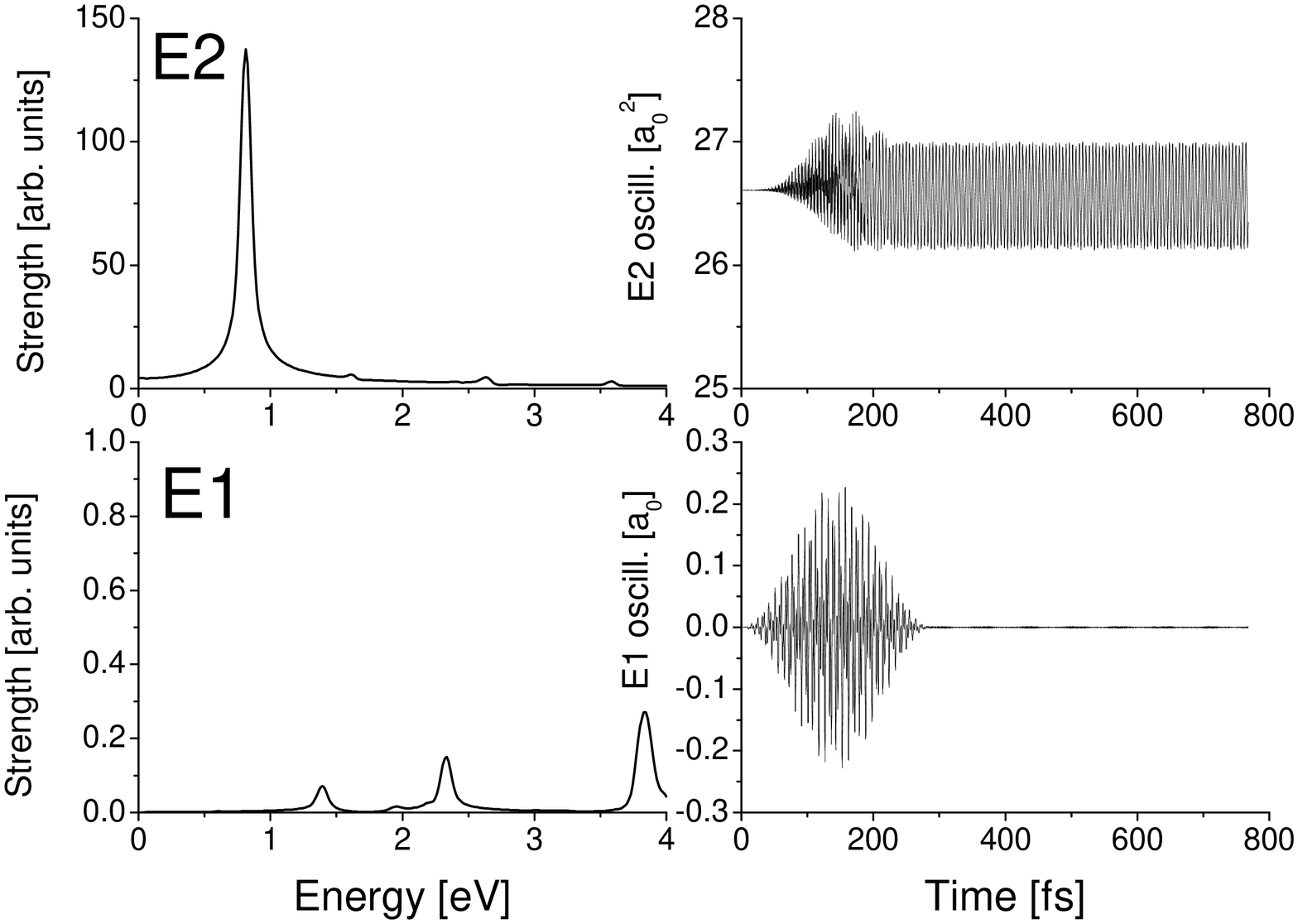}
}
\caption{\label{fig:orsr_iso_fig3}
ORSR in Na$^+_{11}$ via the isolated dipole state at 1.35 eV.
The left panels show quadrupole and dipole strengths as function of
the excitation energy.  The right panels display the electronic
quadrupole and dipole moments (in atomic units) as a function of
time. It is seen that, while the quadrupole oscillation is persistent,
(right-top plot), the dipole signal exists only during the coinciding
pump and Stokes pulses (right-bottom plot).  The calculations were
performed for laser frequencies $\hbar \omega_{s}=$0.46 eV and $\hbar
\omega_{p}=$1.27 eV, intensities $I_{s}=1.5 I_{p}= 2.2 \cdot 10^{10}
W/cm^2$ and pulse durations $T_{s}=T_{p}= 300$ fs. The detuning from
the intermediate state is $\Delta \sim $0.08 eV.  }
\end{figure}

\section{Results and discussion}
\subsection{ORSR for low-frequency isolated quadrupole}

Figure \ref{fig:orsr_iso_fig3} shows the ORSR via the isolated dipole
state at 1.35 eV with a detuning of $\Delta \sim $0.08 eV.  The right
panels show time evolution of the dipole and quadrupole moments.  It
is seen that the ORSR mechanism leads to {\it enduring} quadrupole
oscillation.  Since electron-ion and electron-electron relaxations are
not taken into account here, these oscillations persist for several ps
and further. The dipole oscillations, on the other hand, exist only
during the pulses at $t=0\!-\!300$ fs. The left panels display the
corresponding dipole and quadrupole strengths in the frequency domain,
obtained as the Fourier transforms of the oscillating moments. The
quadrupole mode of interest at 0.81 eV dominates all other quadrupole
excitations, even the quadrupole plasmon. So, just this mode is presented
in the enduring oscillation seen in the right-top panel.  The dipole
strength is negligible (compare different scales of the top and bottom
panels) and so should not noticeably compete with the quadrupole mode
of interest. As was shown in \cite{Ne_PRA_2006}, a significant decoupling of
competing modes is crucial for detection of the target quadrupole
state.

%
\begin{figure}
\centerline{
\includegraphics[height=11cm,width=8cm,angle=-90]{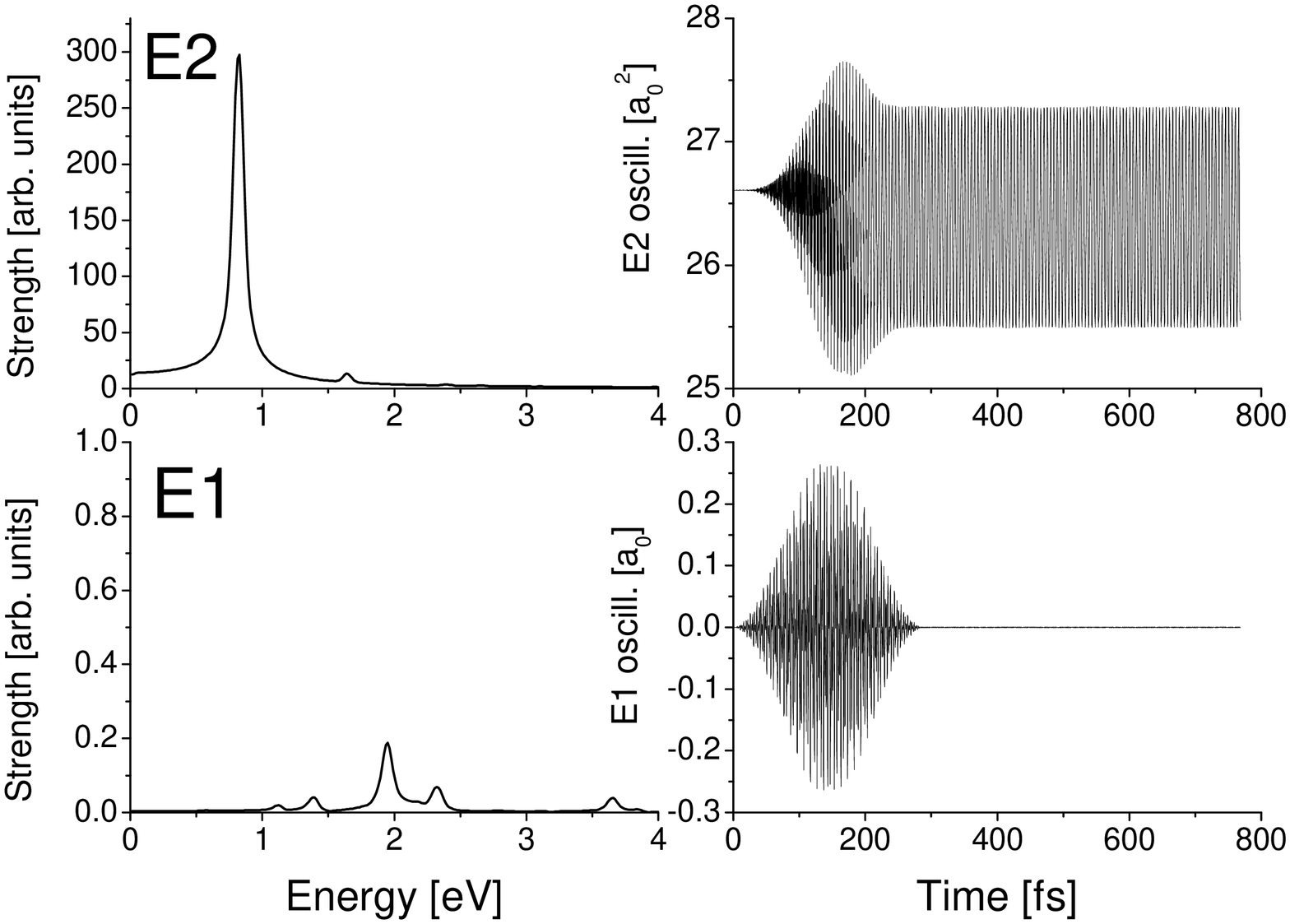}
}
\caption{\label{fig:orsr_pla_fig4}
Similar as figure 3 but for an intermediate dipole frequency of
2.05 eV. The calculations were performed for the laser frequencies
$\hbar \omega_{s}=$1.24 eV and $\hbar \omega_{p}=$2.05 eV, intensities
$I_{s}=1.5 I_{p}= 1.44 \cdot 10^{10} W/cm^2$ and pulse durations
$T_{s}=T_{p}= 300$ fs. The detuning is $\Delta \sim$0.15 eV with
respect to the dipole state at 1.9 eV and $\Delta \sim$0.25 eV with
respect to the dipole plasmon at 2.3 eV.  }
\end{figure}
Figure \ref{fig:orsr_pla_fig4} shows results for the ORSR via the
region between two close dipole structures. The pump laser frequency
2.05 eV is placed between the isolated peak $1eh$ at 1.9 eV
and the tail of the dipole plasmon lying at 2.3 eV, thus representing a
considerable detuning from both dipole structures. As these structures
are much stronger than the intermediate dipole in the previous case,
it becomes possible to get twice larger quadrupole signal even at the
lower laser intensity ($I_{s}=1.5 I_{p}= 1.44 \cdot 10^{10} W/cm^2$
against $I_{s}=1.5 I_{p}= 2.2 \cdot 10^{10} W/cm^2$ in
figure \ref{fig:orsr_iso_fig3}).  Like in the previous case, we have no
appreciable competitors though now the pump frequency is rather close
to the dipole plasmon.  Altogether, figures \ref{fig:orsr_iso_fig3} and
\ref{fig:orsr_pla_fig4} justify that one can get robust ORSR signals
in clusters via both isolated dipole states and tail of the
dipole plasmon.

One may observe in the right-up panel of figure \ref{fig:orsr_pla_fig4}
that the quadrupole oscillation leads to some shift of
the average quadrupole moment. Indeed, the oscillation starts at the moment
26.6 $a_0^2$ but then proceeds around a bit lower average value $\sim 26.4 \; a_0^2$.
Our analysis shows that this effect is caused by a non-isotropic emission of
electrons from the cluster. Indeed, the axial cluster Na$^{+}_{11}$
has a shape of the prolate ellipsoid. The quadrupole oscillation of multipolarity
$\lambda\mu=20$ drives electrons along the symmetry axis of the cluster and thus favors
emission of electrons from the poles of the cluster ellipsoid. This makes
the shape of the electron subsystem more spherical and therefore
effectively decreases the quadrupole moment. The stronger the oscillation, the
larger the moment shift. Hence the effect is most apparent in figure \ref{fig:orsr_pla_fig4}
which exhibits the strongest quadrupole mode. However, even in this case the moment shift
is quite modest (~2\%) and thus should not noticeably influence the accuracy
of measurements of the energy of the quadrupole mode in ORSR experiments.

\subsection{Coherence and population for the target state}

%
\begin{figure}[t]
\centerline{
\includegraphics[height=8cm,width=7cm,angle=-90]{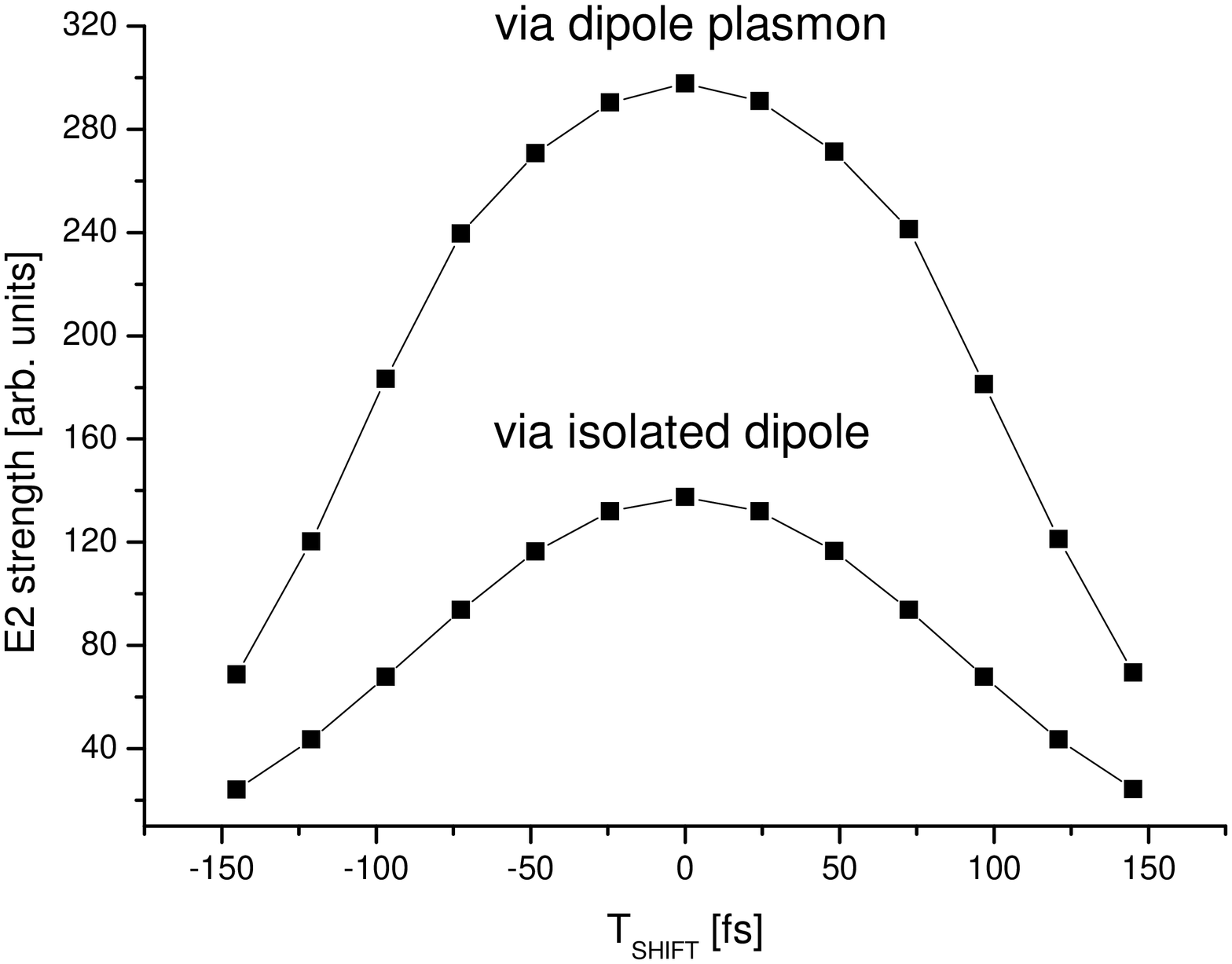}
}
\caption{
\label{fig:fig5}
The quadrupole strength as a function of the pulse shift for ORSR
via the isolated dipole state  and the dipole plasmon.
}
\end{figure}
Figures \ref{fig:orsr_iso_fig3} and \ref{fig:orsr_pla_fig4} were obtained
with simultaneously active pump and Stokes pulses, i.e. with the relative
time shift $T_{\rm shift}=0$. The dependence of the quadrupole peak
height on the time shift it shown in figure \ref{fig:fig5}.  It is
seen that the maximal quadrupole strength is achieved at coincident
pump and Stokes pulses. A similar picture emerges for stimulated Raman
scattering, when plotting the population of the state instead of its strength
\cite{Scoles,SEP,Vitanov}. However, such a correlation between the
strength and population arises only at low population and does not
imply equality of these two characteristics. This point deserves a closer
inspection.

Let us confine the formal considerations to the active subspace.  At
any time $t$, the many-body wave function of the three-level $\Lambda$
system can be represented as a superposition
\begin{equation}
\psi(t)=c_0(t)|0\rangle+c_1(t)|1\rangle+c_2(t)|2\rangle
\end{equation}
where $c_0(t)$, $c_1(t)$ and $c_2(t)$ are time-dependent amplitudes of
the initial, intermediate and final bare states, respectively. The
population of the target quadrupole state then reads $|c_2(t)|^2$. But
the strength of the quadrupole transition (considered in figures
\ref{fig:orsr_iso_fig3}-\ref{fig:fig5}) is
$\sim c_0(t)*c_2(t)\langle2|E2|0\rangle^2$ and so corresponds not to
the population but to the coherence $c_0(t)*c_2(t)$ of the initial and
target states.  The population and coherence have different
behaviors. They both grow at the onset of the population of the level
$|2\rangle$ but then the coherence gets the maximum at
$c_0^2(t)=c_2^2(t)=0.5$ and begins to vanish with further increasing
$|c_2(t)|^2$ from 0.5 to 1.

The calculation of the population $|c_2(t)|^2$ in TDLDA is somewhat
involved. So, in the present study, we will only consider a simple
estimate. We know that the dominant component of the low-frequency
quadrupole state is the $1eh$ configuration $[200]_e[220]_h$ (see
Table 1). To populate this configuration, the electron from the
occupied state $[220]$ should be transferred to the unoccupied state
$[200]$.  This should manifest itself in the time dependent
single-particle wave function $\phi_{[220]}(\vec r,t)$. Namely, it has
to coincide with the initial state $\bar{\phi}_{[220]}(\vec r)$ at t=0
and then acquire large contribution of $\bar{\phi}_{[200]}(\vec r)$ in
the course of time. Hence the population $P^{eh}_2(t)$ of the
$[200]_e[220]_h$ quadrupole component can be estimated as the squared
overlap
\begin{equation}
 P^{eh}_2(t)= |\int d\vec r \phi_{[220]}(\vec r,t) \bar{\phi}_{[200]}(\vec r)|^2 .
\end{equation}
This estimate yields for $t>600$ fs (i.e. for the time when, following
figures \ref{fig:orsr_iso_fig3} and \ref{fig:orsr_pla_fig4}, only the
low-lying quadrupole mode survives) quite
small population 0.01-0.03.  Instead, the overlap with the initial static state
\begin{equation}
 P^{in}_2(t)= |\int d\vec r \phi_{[220]}(\vec r,t) \bar{\phi}_{[220]}(\vec r)|^2
\end{equation}
turns out to contribute the complementing 99-97\%. Similar relations
were obtained for the direct two-photon (DTP) excitation of the
quadrupole, described in \cite{Ne_PRA_2006}.  During the time
evolution, the single-electron wave functions thus mainly keep their
initial structure and only a small fraction (a few per cent) of the
intended $1eh$ quadrupole configuration $[200]_e[220]_h$ is really
populated. This is mainly the consequence of using the
large detuning. One might be tempted to decreasing the detuning
or, alternatively, enhance the population by increasing the laser
intensity. But then we would run into the on-linear regime of TDLDA where
the cross coupling between numerous states of the system takes place
and hence the three-level picture most probably fails.
This trouble reflects  the more complex dynamics of metal clusters
as compared to simple molecules. In any case, even
the population of a few percent has to be sufficient to detect the quadrupole
state in experiment. The TDLDA calculations for the DTP in
Na$^+_{11}$ showed measurable signatures of the low-lying quadrupole
state in PES \cite{Ne_PRA_2006}. A similar situation is expected for
the ORSR.

\subsection{ORSR stability}

%
\begin{figure}
\centerline{
\includegraphics[height=14cm,width=9cm,angle=-90]{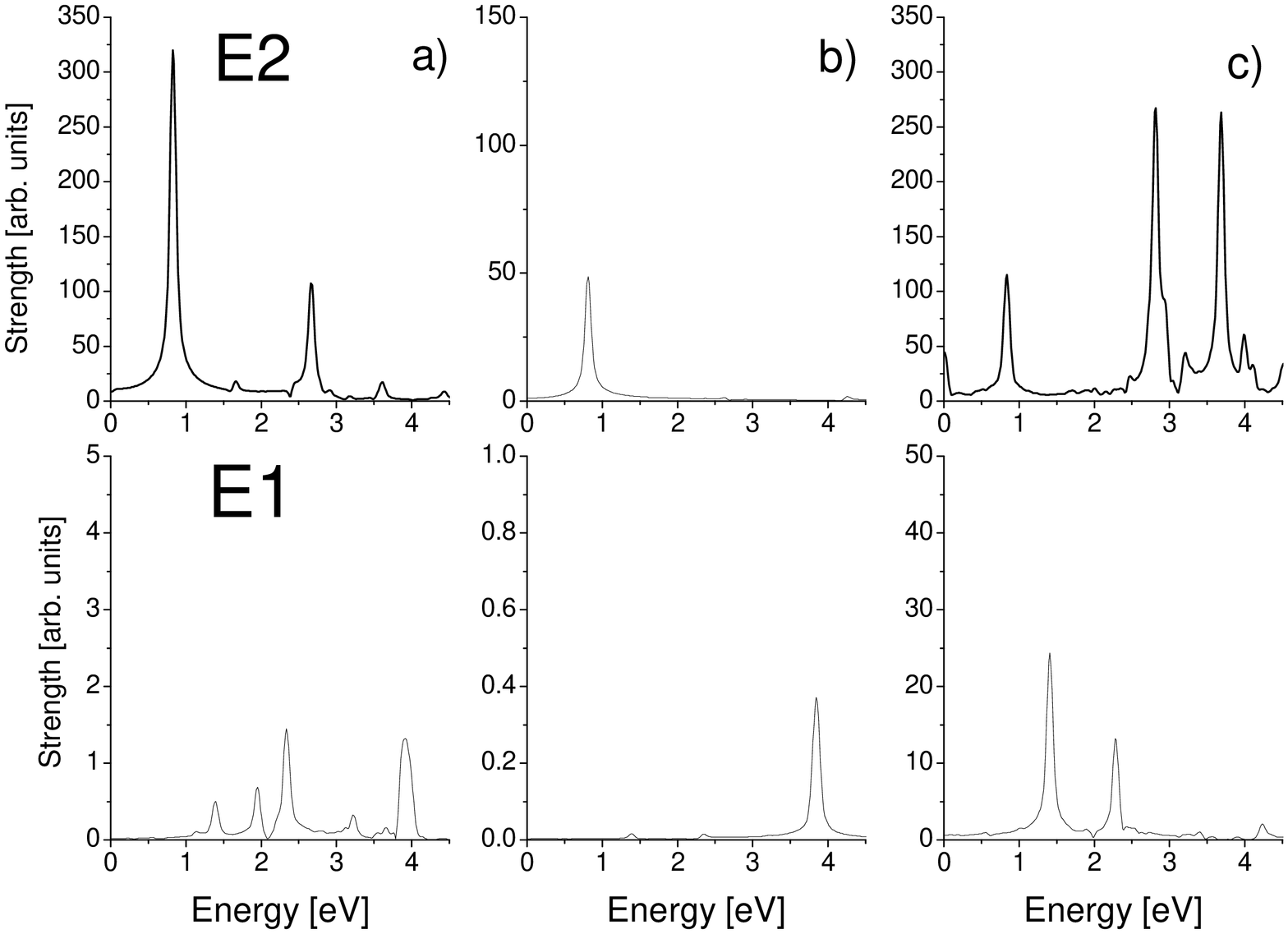}
}
\caption{
\label{fig:fig6}
The strengths calculated as in figure \protect\ref{fig:orsr_iso_fig3} but
a) for larger intensities $I_{s}=1.5 I_{p}=7.44 \cdot 10^{10} W/cm^2$;
b) with small deviation 0.03 eV from the two-photon resonance;
c) without detuning from the intermediate dipole state. See text for
more details.
}
\end{figure}
It is still necessary to check the stability of the ORSR scheme to
variations of the process parameters.  Our tests are illustrated in
figure \ref{fig:fig6}. It shows ORSR via the isolated dipole. The
process via the tail of the dipole plasmon (not shown here) produces
the same features.

The panels a) of figure \ref{fig:fig6} show the quadrupole and dipole
strengths at larger pulse intensities. The intensities are increased
by a factor 3.4 as compared to figure \ref{fig:orsr_iso_fig3}.  This
allows to get a twice stronger quadrupole mode at 0.81 eV. However, we
are punished by stronger dipole excitations and a coupling to the
quadrupole mode at 2.8 eV (as a part of the quadrupole plasmon), which
can complicate discrimination of the target state in PES.  So, though
the ORSR works even at high intensities, the lower intensities are
better suited for the experimental analysis. On the other hand, it is not
worth to go below the optimal intensities of figure \ref{fig:orsr_iso_fig3}
since this would
lead to an unnecessary weakening the quadrupole strength.

The panels b) in figure \ref{fig:fig6} show that small deviations from
the two-photon resonance condition $\omega_p-\omega_s=\omega_2$ lead
to weakening of the target mode. Nevertheless, because of the finite
width of the mode, the signal does not vanish too rapidly.  In fact
the width of the mode determines the maximal deviation for $\omega_s$
while looking for the mode in the experiment.

Finally, the panels c) of the figure demonstrate what
happens without detuning from the intermediate dipole state. In this
case, the dipole strength of the intermediate state is considerably
increased while the population of the target quadrupole state
shrinks. Besides that, a large fraction of competing high-frequency
quadrupoles appear. So, a considerable detuning is crucial for the
success of the ORSR scheme.

\subsection{DTP exploration of quadrupole plasmon}

The top panel of figure \ref{fig:fig6}c deserves a deeper analysis because
it demonstrates an important feature of ORSR and similar two-photon
processes in metal clusters.  Indeed, if two photons from the pump and/or
Stokes pulses happen to come into resonance with one of the peaks of the
quadrupole plasmon, then this peak is strongly excited. Such a case is
observed in the top panel of figure \ref{fig:fig6}c). Comparison of
this plot with Fig. 2 shows that $2\hbar\omega_p = 2.7$ eV and
$2\hbar\omega_p + 2\omega_s = 3.7$ eV: this approximately covers two
of the plasmon states and thus result in two prominent pikes at these
energies.
Obviously, such undesirable excitation of the quadrupole plasmon can
spoil the discrimination of low-frequency quadrupole modes in PES.  At
the same time, this provides new possibilities for
simultaneous investigation of low- and high-frequency quadrupoles.

%
\begin{figure}[t]
\centerline{
\includegraphics[height=11.0cm,width=9.5cm,angle=-90]{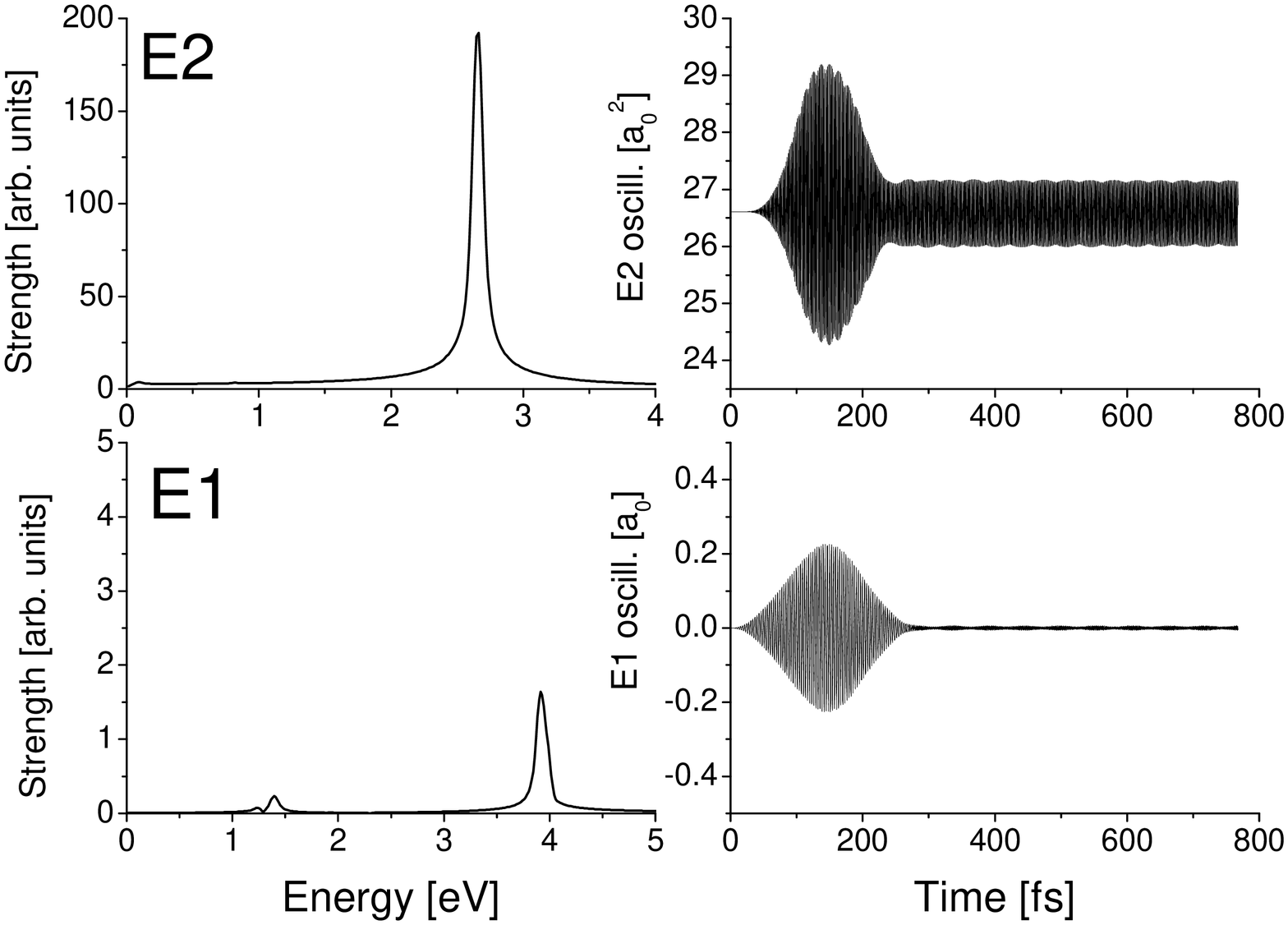}
}
\caption{
\label{fig:fig7}
Direct two-photon excitation of the quadrupole plasmon state at 2.6 eV
in Na$^+_{11}$.
The cluster is irradiated by the pump laser only, with characteristics
$\hbar \omega_{p}=$1.31 eV, $I_{p}= 2.2 \cdot 10^{10} W/cm^2$ and
$T_{s}=300$ fs.
}
\end{figure}

The example above indicates that implementation of two lasers with
different frequencies complicates the analysis of the quadrupole
plasmon because we may have at once more than one resonance.
This would hinder the detection procedure. For the same reason
is not optimal to use for investigation of high-frequency states
the ORSR in the ladder configuration.
It is much better to apply the DTP method proposed in
\cite{Ne_PRA_2006}, where the quadrupole state of interest is
populated by absorption of two photons from one laser, as shown in
figures \ref{fig:lam_sys_fig1} and \ref{fig:be12_fig2}.
An example of such a DTP process is presented in figure \ref{fig:fig7}
where the resonant absorption of two photons from the pump laser
results in a strong excitation of the quadrupole plasmon state at
2.6 eV. Unlike the top panel of Fig. 6c), only one frequency is
involved and thus only one resonant peak excited. A properly tuned DTP
with scanning frequencies allows to investigate the quadrupole plasmon
state by state.

\subsection{Proposed experiment}

The discussion above shows that the most optimal way is a
simultaneous investigation of the low- and high-frequency quadrupole
states by a combination of ORSR and DTP methods, respectively.
For this aim we need three synchronized lasers: tunable infrared pump
and Stokes lasers and ultraviolet probe laser.

The detection scheme can be the same as proposed in \cite{Ne_PRA_2006}.
Namely,
the cluster is irradiated  by a probe pulse (with an appropriate delay)
leading to a direct emission of electron out of the excited quadrupole state.
The coupling of the quadrupole oscillation to the single-electron PES
structures will create  the satellites in the PES.
Thus, by recording the PES and measuring the relative frequencies of
the satellites, one may determine the frequency of the quadrupole state.

The experiment should follow three steps.

1) As a first step, we should find the optimal parameters
(intensity, duration, ...) for the probe pulse responsible
for the photoionization of the cluster and detection process.
For this aim we should scan the pulse parameters so as to get
the strongest and, at the same time, distinct single-particle PES
from the ground state. Our predictions \cite{Ne_PRA_2006}
for the optimal pulse intensities and durations
($I =2\cdot 10^{10} - 2\cdot 10^{11} W/cm^2$ and $T =
200 - 500$ fs) can be used here as a first guess. These predictions
are relevant for all three pulses (pump, Stokes and probe)
which may have similar characteristics, apart from
the photon frequency.

In principle, the photoionization can be also provided by the coherent
synchrotron source with high-frequency photons.  Then the two-photon ionization
proposed in \cite{Ne_PRA_2006} can be replaced by more effective
one-photon ionization. In this case, the characteristics of the pump and
Stokes pulses can be taken again from \cite{Ne_PRA_2006}, while the parameters
of the probe irradiation need an independent adjusting.

2) Now we can proceed with the next step: to explore the high-frequency
states of the quadrupole plasmon. We use here only the pump and probe
lasers. The pump frequency is scanned until its double value comes into
resonance with the states of the quadrupole plasmon. The probe pulse
should have sufficient delay with respect to the pump pulse so as to be
safely decoupled from it and to measure self-sustaining oscillations only.
Since the quadrupole
plasmon has high energy, one-photon ionization by a probe laser in the
visible range suffices for our aims. The maximum satellite
signal provides the quadrupole energy in two ways, first as the double
pump frequency and second (as a countercheck) from the offset of the
satellites.  Thus one can explore, state by state, all the quadrupole
plasmon \cite{comm2}.

3) In a final step, one can refine the measurement of the low-frequency
quadrupole state.  First, one should choose the optimal intermediate
dipole frequency by combining a suitable detuning above or below the
dipole state with the feature that the double pump frequency lies
between the peaks of the
quadrupole plasmon and thus avoid their resonance excitation. By such
a way we will hopefully minimize the influence of the quadrupole
plasmon. Then one should scan the Stokes laser until the two-photon
resonance $\omega_p-\omega_s=\omega_2$ with a low-frequency quadrupole
is achieved.

The proposed scheme allows to obtain not only the frequencies of the
quadrupole states but also their lifetime. To that end, one
should simply increase, step by step, the delay between the pump and
probe pulses. The relaxation of quadrupole oscillation will finally
lead to an extinction of the satellites from which one can read off
the lifetime.

\section{Conclusions}

In this paper, we have proposed a combined exploration of electronic
quadrupole low-frequency (infrared region) and high-frequency (region of the
quadrupole plasmon) states in free
metal clusters by means of two-photon processes: direct
two-photon (DTP) excitation and off-resonant stimulated Raman (ORSR)
scattering. The DTP uses one pump laser and retrieves the two photons from
it while the ORSR employs a pump and a Stokes pulses with different
frequencies. The final proof of the successful excitation of a quadrupole
mode is achieved by a probe pulse with a subsequent measurement of the
photo-electron spectra (PES). The present analysis
is based on realistic simulations within the time-dependent local density
approximation. The main attention is paid to the ORSR which, for our knowledge,
still have never been considered for atomic clusters.

The calculations show that the high-frequency quadrupole states in the
regime of the quadrupole plasmon are preferably explored by DTP.  The
more flexible ORSR is a powerful tool to investigate the isolated
low-frequency (infrared) quadrupoles. ORSR allows to use various
intermediate dipole states, from the isolated infrared dipoles to
the dipole plasmon. In all the cases, an appreciable
detuning from true dipole states is crucial to avoid unwanted cross
talk. A proper combination of both methods (DTP, ORSR) should allow to explore
both spectra and lifetimes of the quadrupole states.  We have worked out
optimal parameters of DTP and ORSR schemes and checked the sensitivity
of the schemes to parameter variations. The proposed two-photon schemes
are quite general and, in principle, can be used for a variety
of clusters including supported and embedded ones.

The low- and high-frequency quadrupole states can deliver important
information on electron-hole excitations inside the valence shell
($\Delta N =0$) and through two shells $(\Delta N =2)$. By combining
the two-photon and photoelectron data, one can get the single-electron
energies above the Fermi level and thus greatly enlarge our knowledge
on the mean field spectra of valence electrons. These spectra give
access to other cluster features (mean field, deformation) and
provide a critical test for the theory of cluster structure.

\begin{acknowledgments}
V.O.N. thanks Profs. K. Bergmann and B.W. Shore for valuable discussions.
The work was partly supported  by the DFG grant  GZ:436 RUS 17/104/05, the grant of
University of Paul Sabatier (Toulouse, France) and
Heisenberg-Landau (Germany-BLTP JINR) grants for 2005 and  2006 years.
\end{acknowledgments}


\begin{thebibliography}{100}
\bibitem{comm1}
  In principle, the quadrupole plasmon can be also explored by the ORSR
  in the ladder configuration when the successive levels lie higher
  than the predecessors. However, as is shown in the subsection 3.4
  this option is not optimal.
\bibitem{comm2}
The pump frequency can accidentally coincide with one of the
isolated dipole states below the dipole plasmon and thus result
in the additional PES satellites. However, these satellites can be
easily recognized through a slight shift of the probe frequency,
see \cite{Pohl_01} for more details.
\end{thebibliography}
\end{document}